\documentstyle [prl,multicol,aps,psfig]{revtex}
\begin{document}
\title{Interference effects in resonant magneto-transport}
\author{D. Mozyrsky$^1$ {\rm ,} L. Fedichkin$^2$ {\rm ,}
 S. A. Gurvitz$^3$ {\rm ,} G. P. Berman$^1$\\
$^1$T-13 and CNLS, Los Alamos National Laboratory,
Los Alamos, NM 87545, USA\\
$^2$ Physics Department, Clarkson University, NY 13699-5820, USA\\
$^3$Department of Particle Physics, Weizmann Institute of Science, Rehovot 76100, Israel}
\maketitle
\begin{abstract}
We study non-equilibrium magneto-transport through a single electron transistor or an impurity. We
find that due to spin-flip transitions, generated by the spin-orbit interaction, the spectral
density of the tunneling current fluctuations develops a distinct peak at the frequency of Zeeman
splitting. This mechanism explains modulation in the tunneling current at the Larmor frequency
observed in scanning tunneling microscope (STM) experiments and can be utilized as a detector for
single spin measurement.
\end{abstract}
\hspace{1.5 cm} PACS:  73.50.-h, 73.23.-b, 03.67.Lx.
\begin{multicols}{2}
Problems involving transport through low dimensional structures have received significant
attention in connection with the rapid development of spintronic\cite{r01} and single
electron\cite{r02} devices. Apart from being likely candidates for becoming components of future
electronic integrated circuits, the use of single electron transistors (SET) as charge detectors
have been contemplated in several solid state quantum computing designs\cite{r1}.

In the present work we demonstrate that the spectral density of current fluctuations of a single
electron transistor in the external magnetic field develops a peak at the electron Zeeman frequency
generated by spin-orbit interactions. We attribute such effect to the interference between the spin
up and spin down components of the transmitted current resulting from the spin flips in the
tunneling process.

As a model system we consider a  heterostructure (for example Si/Ge) schematically shown in
Fig.~1. The two regions, to the right and to the left from the dotted line denoting the interface,
have different $g$-factors, $g_1 \approx 2$ for the left region and $g_2 \ne 2$ for the right
region. There are two contacts/Fermi reservoirs in each of the regions. The left region also
contains a quantum dot, so that when a potential difference $V$ is applied between the two
reservoirs, electrons can tunnel from left to the right reservoirs via the dot. The energy levels
of the dot are spin-split by an external magnetic field. In this case the spin-orbit coupling
causes the spin-flip transitions resulting in coherent effects in the tunneling
current\cite{baran}.

We describe our system by the Hamiltonian $H = H_L + H_R + H_S + H_C + H_T$ where the first two
terms represent the unperturbed states of two contacts, $H_L = \sum_{l,s} \epsilon_{l
s}a^{\dag}_{l s} a_{l s}$ and $H_R = \sum_{r,s} \epsilon_{r s} a^{\dag}_{r s} a_{r s}$, where
$a^{\dag}_{ls}$ ($a^{\dag}_{rs}$) creates a fermion/electron at the energy level $\epsilon_l$
($\epsilon_r$) and with spin $s$ in the left (right) reservoir. We assume that there is a single
discrete level in the dot due to spatial quantization. The level is spin-split by the magnetic
field $B$, so that the states in the dot are described by $H_S = \sum_s \epsilon_s {\hat n}_s$,
where $\hat n_s=a^{\dag}_s a_s$, and $a^{\dag}_s$ creates an electron in the dot at the level
$\epsilon_s$ with spin $s$. We denote $\epsilon_{-1/2} - \epsilon_{1/2} = g\beta B \equiv E$,
Fig.~1, where $g$ is the electronic $g$-factor in the dot and $\beta$ is Bohr's magneton. The term
$H_C = \sum_s {U \over 2} {\hat n}_s {\hat n}_{-s}$ corresponds to the Coulomb charging energy for
the electron in the well. In what follows we will assume the case of complete Coulomb blockade,
i.e., $U \rightarrow \infty$, thus allowing for only one electron to occupy the two spin states in
the dot.
\begin{figure}
{\centering{\psfig{figure=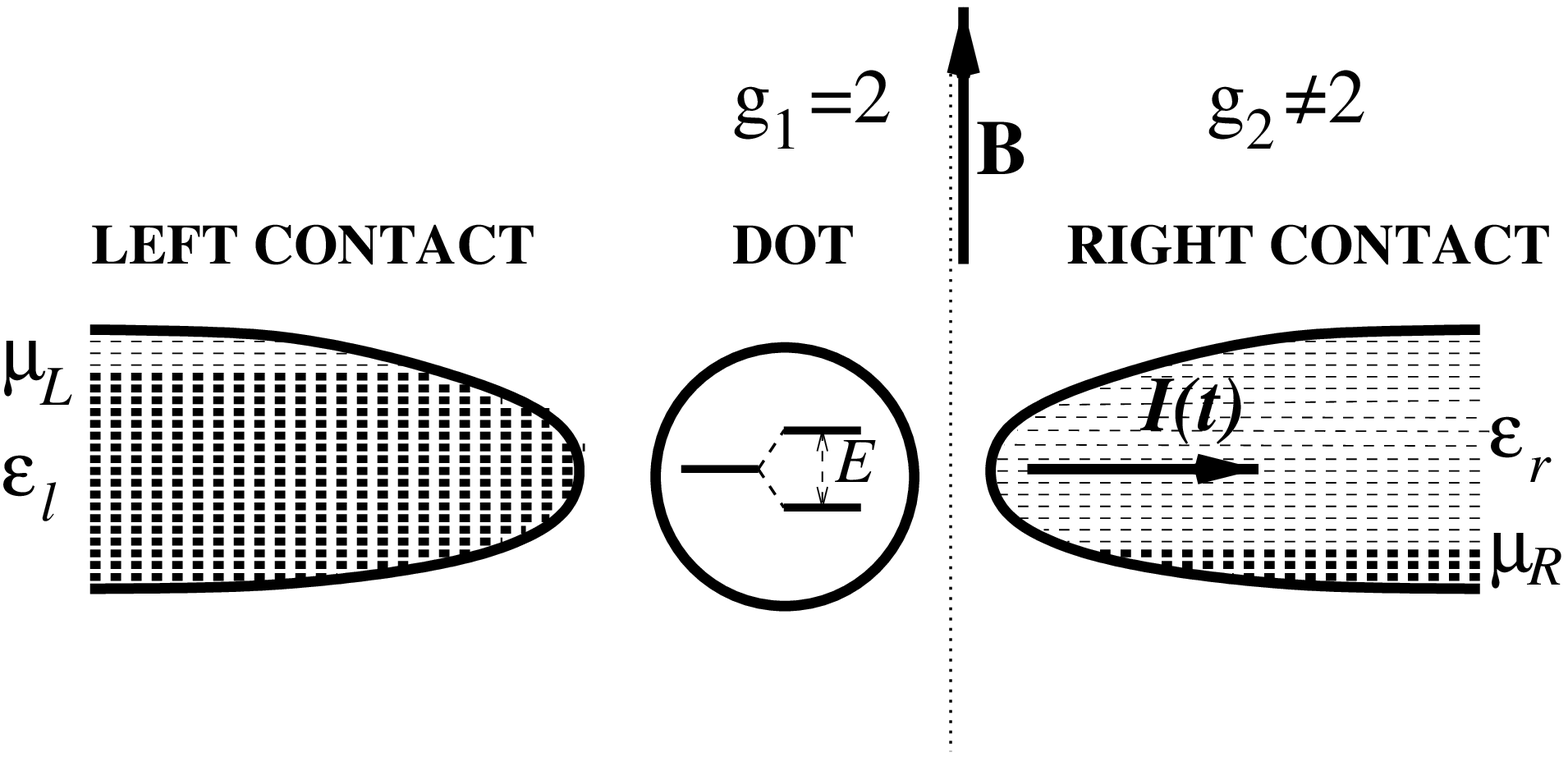,height=4.6cm,width=8.3cm,angle=0}}} {\bf Fig.~1:} Quantum dot
coupled to two contacts. The right contact has $g$-factor different from that of the left contact
and the dot. Tunneling with spin flip generates effective coupling between the two Zeeman sublevels
in the dot.
\end{figure}
The tunneling transitions between the left reservoir, the dot and the right reservoir are
represented by the Hamiltonian:
\begin{eqnarray}
H_T = \sum_{l,s} \Omega_l \left( a^{\dag}_{l s} a_s + a^{\dag}_s a_{l s}
\right)&& \nonumber\\
+  \sum_{r,s,s^{\prime}} \Omega_{r s s^{\prime}}&& \left( a^{\dag}_{r s}a_{s^{\prime}} +
a^{\dag}_{s^{\prime}} a_{r s} \right) \, . \label{a1}
\end{eqnarray}
Here we use gauge in which tunneling amplitudes $\Omega_l$ and $\Omega_{r s s^{\prime}}$ are
real.  As we noted above the key point of our work is that we consider the tunneling transitions
accompanied by spin flips . These are generated by the second term in Eq.~(\ref{a1}) due to
g-factor difference between the dot and the right contact. The mechanism generating such
transitions is similar to that of spin scattering by nonmagnetic impurities in
semiconductors\cite{r9}. Due to spin-orbit interaction, relatively strong in the right contact in
our case, the orbital and spin states of the electron in the right reservoir are mixed, resulting
in effective $g$-factors for the electrons there to be different from $2$. The eigenstates of
$H_R$ are represented by Kramers doublet, $|\psi_{r,s=1/2}\rangle = u_r|\uparrow\rangle +
v_r|\downarrow\rangle$ and a Kramers conjugate state $|\psi_{r,s=-1/2}\rangle$, where $u_r$ and
$v_r$ are functions of spatial coordinates only. We have assumed the spin orbit coupling in the
left reservoir and the dot is much weaker ($g\approx 2$), so that we can neglect by the spin-orbit
mixing effect there. In order to evaluate the tunneling matrix elements for transitions from the
dot to the right reservoir, given by the second term in Eq.~(\ref{a1}), one can utilize Bardeen's
formula\cite{r10}: $\Omega_{r s s^{\prime}} = 1/2m \int d {\vec S}\cdot (\phi_s^{\ast} {\vec
\nabla} \psi_{r, s^{\prime}} - \psi_{r, s^{\prime}} {\vec \nabla} \phi_s^{\ast})$, where the
integral is over any surface lying entirely within the tunneling barrier, separating the dot and
the right reservoir, and the wave functions $\phi_s$ (state with spin $s$ in the dot,
$|\phi_s\rangle = |\phi\rangle|s\rangle$) and $\psi_{r, s^{\prime}}$ are smoothly continued under
the barrier; $m$ is electron's mass and $\hbar = 1$. It is obvious that the states $\psi_s$ under
the barrier are still spin-orbit mixed due to the continuity condition. Therefore the tunneling
matrix elements, corresponding to the transitions from the resonant level to the right reservoir
without spin flip, are $\Omega_{r s s} = 1/2m \int d {\vec S}\cdot (\phi^{\ast} {\vec \nabla} u_r
- u_r {\vec \nabla} \phi^{\ast})$, and the matrix elements of transitions accompanied by spin
flips are $\Omega_{r s \, {\bar s}} = 1/2m \int d {\vec S}\cdot (\phi^{\ast} {\vec \nabla} v_r -
v_r {\vec \nabla} \phi^{\ast})$; ${\bar s}\equiv -s$. For relatively small deviations of $g$
factor in the right reservoir from $2$, $|v| \sim O(|\Delta g u|)$, $\Delta g = g - 2$\cite{r9},
and so the two transition amplitudes are related as $|\Omega_{r s {\bar s}}| \sim O(|\Delta g
\Omega_{r s s}|)$. For $\Delta g > 1$, the two components $u_r$ and $v_r$ are of the same order of
magnitude and so $\Omega_{r s s}\sim\Omega_{r s {\bar s}}$.

In this work we are interested in spectral properties of the tunneling current and calculate its
spectral density. Typically calculations of this sort involve evaluation of the two-particle
Green's functions, which is a quite formidable task in non-equilibrium case, beyond the
applicability of the linear response theory. Instead, we adopt an alternative approach developed
in Refs.\cite{r8,rr8}, that allows one to evaluate the transport rate equations from the
microscopic Hamiltonian. In this letter we show that one can obtain the spectral density of
fluctuations from these equations as well (see below). In the following we outline our calculation
of current spectral density and analyze the obtained results.

We construct the time dependent wave function of the system as
\begin{eqnarray}
|{\bf \Psi}(t)&&\rangle = \Big\{ b_0(t) + \sum_{l,s}[ b_{ls}(t) a^{\dag}_s a_{ls} + b_{l{\bar
s}}(t) a^{\dag}_s a_{l{\bar
s}}]\nonumber\\
&&+ \sum_{l,r,s}[ b_{lrs}(t) a^{\dag}_{rs} a_{ls} + b_{lr{\bar s}}(t) a^{\dag}_{rs} a_{l{\bar
s}}]+ \  ... \ \Big\}|{\bf 0} \rangle\, , \label{a2}
\end{eqnarray}
where the ``ground'' state $|{\bf 0} \rangle$ corresponds to the situation when all states below
Fermi energy in the left contact are filled, while all states above Fermi energy in the right
contact are empty. The above wave function is a superposition of all possible particle-hole
combinations that can be generated by the Hamiltonian $H$; note that $H$ conserves the total
number of particles in the system. Thus the first term in ${\bf \Psi}$ is the amplitude of the
unperturbed state, i.e., when no excitations in the system is present, the second term describes a
state in which a hole is created in the left reservoir and a particle with the same spin occupies
the resonant level, etc. The above wave function satisfies the Schrodinger equation $i|{\bf \dot
\Psi} \rangle = H |{\bf \Psi} \rangle$.

In order to describe transport in our model we introduce probabilities for the dot to be empty or
occupied, provided that a certain number of electrons have passed through the junction. The level
can be either empty, with probability $\sigma_{aa}^{n}$, where the subscript $aa$ indicates that
there is no electrons in the dot and the superscript $n$ describes that $n$ electrons have arrived
in the right reservoir/collector, or the level can be filled with probabilities $\sigma_{bb}^{n}$
and $\sigma_{cc}^{n}$, where $bb$ indicates that the lower Zeeman sublevel $s=1/2$ is filled,
while $cc$ stands for the upper Zeeman sublevel $s=-1/2$ being filled. Occupation of both Zeeman
levels in the dot by two electrons is prohibited in our model by the infinite charging energy $U$;
see Refs.\cite{r8,rr8} for detailed discussion. We also introduce the off-diagonal elements
$\sigma_{bc}^n$ describing coherent superpositions of states on the upper and lower Zeeman levels
of the electron in the dot. $\sigma_{ij}^n$'s are related to the wave function $|{\bf
\Psi}\rangle$ as $\sigma_{aa}^0 = |b_0|^2$, $\sigma_{bb}^0 = \sum_{l,s=1/2} |b_{ls}|^2 +
\sum_{l,s=-1/2} |b_{l{\bar s}}|^2$, $\sigma_{aa}^1 = \sum_{l,r,s} [|b_{lrs}|^2 + |b_{l,r,{\bar
s}}|^2]$, etc.

Following steps of Refs.\cite{r8,rr8} one derives the rate equations for the density matrix
$\sigma$ from the Schrodinger equation for the wave function $|{\bf \Psi} \rangle$. These rate
equations for a general case are presented in\cite{rr8}. One finds for our case:
\begin{mathletters}
\label{a4}
\begin{eqnarray}
&&{\dot \sigma}_{aa}^n = -2\Gamma_L\sigma_{aa}^n + \Gamma_R
\left(\sigma_{bb}^{n-1}+ \sigma_{cc}^{n-1}\right)\nonumber\\
&&~~~~~~~~~~~~~~~~~~~~~~~~~~~~~~~~~~~ + \Delta\Gamma_R \left(\sigma_{bc}^{n-1}
+ \sigma_{cb}^{n-1}\right) \,,\\
\label{aa} &&{\dot \sigma}_{bb}^n = -\Gamma_R\sigma_{bb}^n + \Gamma_L \sigma_{aa}^n -
{\Delta\Gamma_R \over 2} \left(\sigma_{bc}^n +
\sigma_{cb}^n\right)\, ,\\
\label{a4b} &&{\dot \sigma}_{cc}^n= -\Gamma_R\sigma_{cc}^n + \Gamma_L \sigma_{aa}^n -
{\Delta\Gamma_R \over 2} \left(\sigma_{bc}^n + \sigma_{cb}^n\right)
\, ,\\
\label{a4c} &&{\dot \sigma}_{bc}^n= i E \sigma_{bc}^n -\Gamma_R\sigma_{bc}^n - {\Delta\Gamma_R
\over 2} \left(\sigma_{bb}^n + \sigma_{cc}^n\right) \, .\label{a4d}
\end{eqnarray}
\end{mathletters}
Here $\Gamma_{L,R}= 2\pi \Omega_{L,R}^2(\epsilon_s) \rho_{L,R} (\epsilon_s)$ and $\Delta\Gamma_R=
2\pi \Omega_R(\epsilon_s)\delta\Omega_R(\epsilon_s)\rho_R (\epsilon_s)$, where we denote
$\Omega_{r s s}\equiv \Omega_R$, $\Omega_{r s {\bar s}} \equiv \delta\Omega_R$. In derivation of
Eqs.~(\ref{a4}) we assumed that the coupling constants $\Omega$'s and the densities of states
$\rho$'s are weakly dependent on energy, and so $\rho_{L,R} (\epsilon_s) = \rho_{L,R}
(\epsilon_{\bar s})$, $\Omega_{L,R} (\epsilon_s) = \Omega_{L,R}(\epsilon_{\bar s})$ and thus rates\
$\Gamma_{R,L}$ for the electrons tunneling into and out of the dot are independent of energy. We
also assumed that $\Gamma_L ,\Gamma_R \ge \Delta\Gamma_R$ and the bias voltage condition,
$V\gg\Gamma_{L,R}$, which is essential for derivation of Eqs.~(\ref{a4}). We point out that
Eqs.~(\ref{a4}) are derived in the limit of small $\delta\Omega_R$, therefore terms of order
$\delta\Omega_R^2$ and higher are neglected in~(\ref{a4}). One sees from Eqs.~(\ref{a4}) that
similar to Bloch equations the two Zeeman levels in the dot are coupled with each other by the
off-diagonal terms (``coherences'') due to spin-flip transitions through continuum with the rate
$\Delta\Gamma_R$.

By summing Eqs.~(\ref{a4}) over the number of electrons in the right reservoir one obtains the
``standard'' Bloch-type equations for the reduced density matrix $\sigma_{ij} = \sum_n
\sigma_{ij}^n$ with $i,j \equiv a, b, c$. These equations, which  look essentially identical to
Eqs.~(\ref{a4}), describe the state of the resonant level independently of the states of the
reservoirs.

From Eqs.~(\ref{a4}) one can derive the dynamics for the expectation value of the tunneling
current in the right (left) reservoir, $\langle I_{R,L}(t) \rangle =ie\langle{\bf \Psi}(t)|[H,\hat
N_{R,L}]|{\bf \Psi}(t)\rangle $, where $H$ is the total Hamiltonian and $\hat N_{R,L}$ are the
operator of the electron (hole) numbers is the right (left) reservoirs. For instance, by using
$\hat N_{R}=\sum_{r,s}a^\dagger_{rs}a_{rs}$ one easily finds from Eq.~(\ref{a2}) that the average
current in the right reservoir can be written as $\langle I_R(t) \rangle = e \langle {\dot N}_R
(t) \rangle$, where $\langle N_R \rangle = \sum_n n(\sigma_{aa}^n + \sigma_{bb}^n
+\sigma_{cc}^n)$. Using  Eqs.~(\ref{a4}) for $\dot{\sigma}^n$, one can sum over $n$ thus obtaining
\[
\langle I_R(t) \rangle = e\Gamma_R[\sigma_{bb}(t)+\sigma_{cc}(t)]+
e\Delta\Gamma_R[\sigma_{bc}(t)+\sigma_{cb}(t)]\, .\] It is easy to check that the transient
behavior of the average current is an oscillatory one (due to coherence terms $\sim\sigma_{bc}$)
with frequency equal to $E$ and approaching stationary value
\begin{equation}
\langle I(\infty) \rangle = {2 e \Gamma_L\Gamma_R \left(E^2 + \Gamma_R^2 - \Delta\Gamma_R^2\right)
\over \left(2\Gamma_L + \Gamma_R\right) \left(E^2 + \Gamma_R^2\right) - 2\Gamma_R
\Delta\Gamma_R^2} \, . \label{a5}
\end{equation}
Actually one measures a circuit current, given by $\langle I(t) \rangle = \alpha \langle
I_R(t)\rangle +\beta \langle I_L(t)\rangle$ where the coefficients $0\le\alpha ,\beta\le 1$ and
$\alpha +\beta =1$ are depending on a circuit geometry\cite{r11}. Obviously, the stationary
current, $\langle I(\infty ) \rangle$  is independent of $\alpha$ and $\beta$, Eq.~(\ref{a5}).
Yet, the transient current, $\langle I(t ) \rangle$ and so the current spectral density are
depending on a circuit geometry. For simplicity we consider such a case where only the collector
current is measured ($\alpha =1, \beta =0$).

In order to evaluate current spectral density , $S_I (\omega) = \int_0^\infty d\tau
\cos(\omega\tau) \langle I(t) I(t+\tau) \rangle$, from rate equations ~(\ref{a4}), we utilize
MacDonald's formula, that relates $S_I$ to the dispersion of charge accumulated on the collector
(right reservoir)\cite{r12}:
\begin{equation}
S_I (\omega) = e^2\omega / \pi \int_0^\infty dt \sin (\omega t) \langle {\dot N}_R^2 (t) \rangle\,
. \label{md}
\end{equation}
The dispersion for the number of electrons in the right reservoir can be found from
Eqs.~(\ref{a4}) as $\langle N_R^2 \rangle = \sum_n n^2 (\sigma_{aa}^n +\sigma_{bb}^n +
\sigma_{cc}^n)$.  The later must be averaged with respect to the stationary state distribution in
the dot. Evaluation of the above sum is tedious but can be performed exactly from Eqs.~(\ref{a4}).
Here we quote the final expression for spectral density $S_I(\omega)$. The general result is
rather cumbersome. In the region of interest, $E \ge \Gamma_L, \Gamma_R \ge \Delta\Gamma_R$,
expanding $S_I$ in powers of $\Delta\Gamma_R$ up to $O(\Delta\Gamma_R^2)$, we obtain:
\begin{eqnarray}
&&S_I (\omega) = {e^2 \over \pi} {2\Gamma_L\Gamma_R \over 2\Gamma_L + \Gamma_R}{4\Gamma_L^2 +
\Gamma_R^2 + \omega^2 \over \left(2\Gamma_L
+ \Gamma_R \right)^2 + \omega^2}\nonumber\\
&&~~~~~~~~~~~~~~~~~~ +{8 e^2 \Gamma_L^3 \Gamma_R^2 \over \pi E^2 \left(2\Gamma_L +
\Gamma_R\right)^2}{\Delta\Gamma_R^2 \over \Gamma_R^2 + \left(\omega - E\right)^2} \, . \label{a11}
\end{eqnarray}
The spectrum~(\ref{a11}) is shown in Fig. 2. The first term in~(\ref{a11}) is the shot noise
approaching the ``Schottky'' limit $S_I = e\langle I \rangle /\pi$ for $\omega \gg \Gamma_R,
\Gamma_L$. For frequencies $\omega \le \Gamma_L, \Gamma_R$ there is a dip in the spectrum - the
result merely consistent with Refs.\cite{r6}. The second term, representing a distinct peak, arises
due to spin-flip transitions between the Zeeman-split sublevels in the dot. It is roughly of
Lorentzian shape centered approximately at $\omega = E$ and having width $\Gamma_R$. Importantly,
the width of the peak is governed by $\Gamma_R$, while the width of the dip is controlled by
$\Gamma_L$ for $\Gamma_L\gg\Gamma_R$. Such condition guarantees that the peak is sufficiently
distinct and thus can be resolved. A similar situation takes place in case of a current tunneling
through a double well structure\cite{r7}, where a peak in the fluctuation spectrum appears to be
located at the tunneling frequency for the double well structure.

The ratio of the peak's height to the noise pedestal (the signal to noise ratio) given by
Eq.~(\ref{a11}) is $S/N = 4\Gamma_L^2 \Delta\Gamma_R^2 / E^2 \Gamma_R (2\Gamma_L + \Gamma_R)$. The
$S$ to $N$ ratio can be significantly increased in heterostructures with greater $g_1 - g_2$
difference, and thus greater spin transition rate $\Delta\Gamma_R$, or in asymmetric SETs with
$\Gamma_L \gg\Gamma_R$.
\begin{figure}
{\centering{\psfig{figure=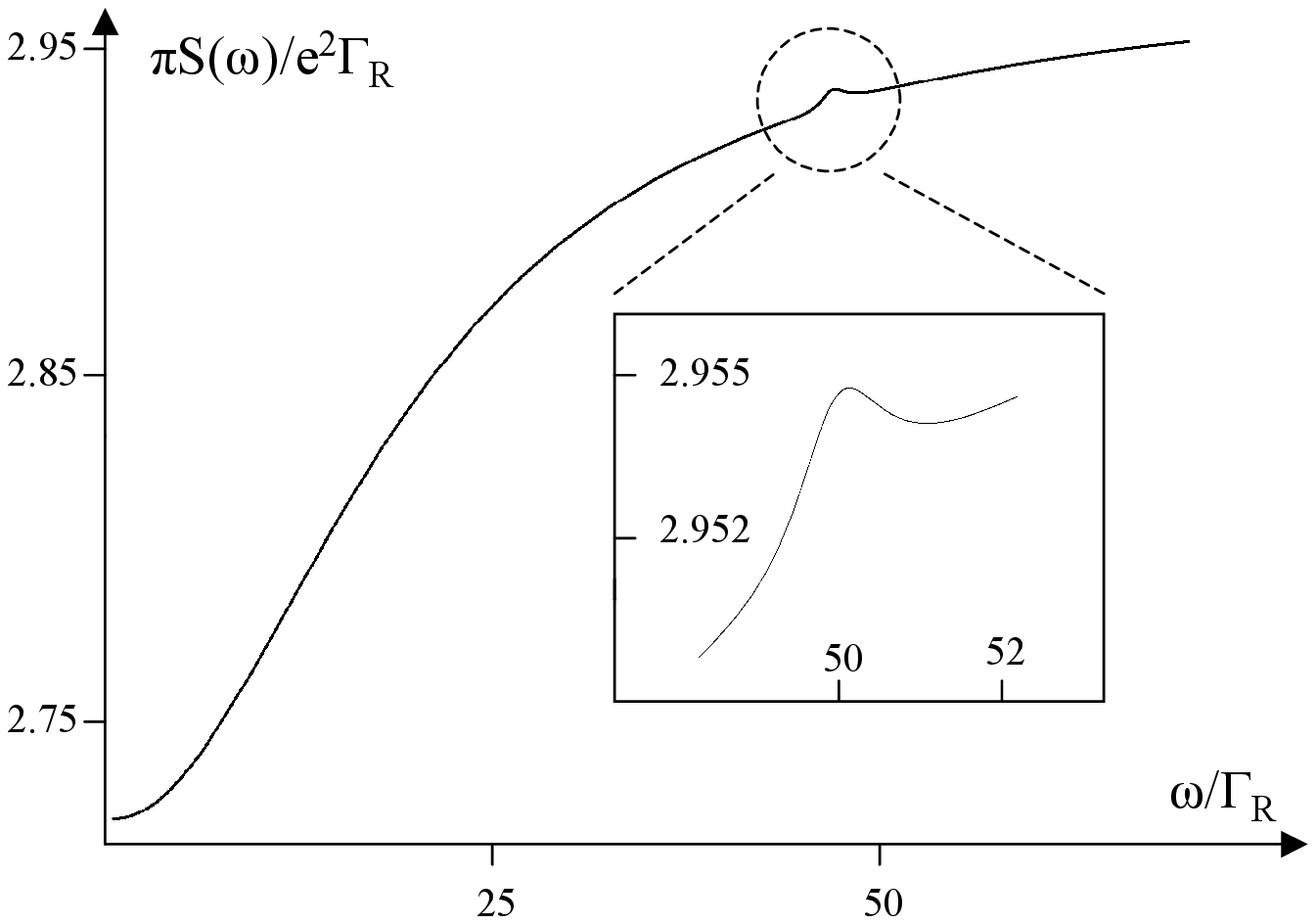,%
bbllx=200pt,bblly=345pt,bburx=590pt,bbury=615pt,%
silent=,width=8.3cm,angle=0,clip=}}} {\bf Fig.~2:}   Power spectrum of tunneling current
fluctuations; Eq.~(\ref{a11}).  Here $E = 50 \Gamma_R$, $\Delta\Gamma_R=0.4\Gamma_R$, $\Gamma_L =
20 \Gamma_R$.
\end{figure}
The above described spin-coherence mechanism can be used for single spin detection. Suppose that
nuclear spins in the dot are polarized. From Eqs.~(\ref{a5}), (\ref{a11}) one can evaluate the
orders of magnitude for parameters needed for observation of the distinct peak in the fluctuations
spectrum in magnetic field generated by a few nuclear spins. The width of the peak in~(\ref{a11})
is defined by the value of current through the structure, $\Gamma_R \simeq \langle I \rangle /e$
for $\Gamma_L > \Gamma_R$. Therefore in order to resolve a peak due to spin flip transitions one
needs to satisfy condition $E>\Gamma_R$, though $E$ should not be too great as signal to noise
ratio decreases with growth of $E$. Assuming that the Zeeman splitting $E$ is solely due to
hyperfine coupling, which is typically of order $10^2$ MHz per single nuclear spin, the measurable
tunneling current through the structure would be of order 100 pA. This number is well within the
capabilities of today's single-electronics. We thus conclude that the interference effect in
resonant magneto-transport can be used for detection of polarization produced by $\le$ 10 nuclear
spins. The sensitivity of such measurements might be higher than that of any existing experimental
setup \cite{r101,r102}.

We argue that the effect considered in the present work can explain coherent oscillations with
Zeeman frequency in the tunneling current, which have been observed in a set of scanning tunneling
microscope (STM) experiments\cite{r5}. Impurities at semiconductor surfaces are known to form
resonant levels that can influence the STM tunnel current~\cite{r6}. In contrast to previous
attempts to explain the experiments\cite{r5}, we suggest that it is not the impurity spin but the
current itself generates coherent oscillations due to the tunneling of electrons with spin flip
via the resonant level formed by the impurity.

We must note, however, that the above condition $E>\Gamma_R$ for the observation of a narrow peak
in the spectral density requires the resonant current ($I_r/e$) to be less than the oscillation
frequency $E$. On the first sight this condition seem to contradict to our statement that the above
discussed interference mechanism can naturally explain experiments~\cite{r5}. Indeed, this peak
has been detected in the opposite regime, namely $I/e
> E$. Yet, in the experiments~\cite{r5} the tunnel current flows through a cluster of impurities
having a number of resonant levels. As a result the current through a single resonant level ($I_r$)
represents only a fraction of the total STM current and therefore is considerably smaller than the
total STM current $I$\cite{r100}. Thus the width of the peak in the noise spectrum in
experiments~\cite{r5} must be defined by a current through a single resonant level rather than by
the total STM current. Therefore we argue that the condition $I/e
> E$ can be met with no contradiction to the principal conclusions of this work.

Finally we emphasize that the noise spectrum for the noise in the circuit current ($I_c = \alpha
I_R + \beta I_L$) can also be calculated using approach developed in this work. Using charge
conservation, $I_L-I_R = \dot{Q}$, where $Q$ is charge in the dot, one obtains a simple relation
for the noise spectra of the tunnel currents through right and left junctions: $S_{I_c}(\omega) =
\alpha S_{I_L}(\omega) + \beta S_{I_R}(\omega) - \alpha\beta \omega^2 S_Q (\omega)$. The spectrum
of charge fluctuations in the dot, $S_Q$, can be derived from the rate equations~(\ref{a4}) by
calculating stationary charge auto-correlation function, say
$\langle\sigma_{aa}(0)\sigma_{aa}(t)\rangle$. This calculation, however, will alter the results
only quantitatively and thus will be rendered to future work\cite{r100}.

The authors acknowledge valuable discussions with H. U. Baranger, J. Brown, M. Hawley, Sh. Kogan,
A. Korotkov, Y. Manassen, V. Privman and I. D. Vagner. This work was supported by the US DOE under
contract W-7405-ENG-36, by the NSA and by ARDA, contracts DAAD 19-99-1-0342 and DAAD 19-02-1-0035.
D.M. and L.F. were supported, in part, by the US NSF grants ECS-0102500 and DMR-0121146.

\end{multicols}
\end{document}